\documentstyle[namedreferences,psfig]{spackapG}
\begin{opening}
\title{On the reduction and presentation of data in astronomical two-channel 
photopolarimetry.} 
\author{G. \surname{Leyshon}}
\institute{\hspace{2.5cm}Department of Physics and Astronomy,\\
\hspace{2.5cm}University of Wales, Cardiff\\
\hspace{2.5cm}P.O. Box 913\\
\hspace{2.5cm}Cardiff, UK\\
\hspace{2.5cm}CF2 3YB\\
\hspace{2.5cm} \\
\hspace{2.5cm}E-mail address: {\tt G.Leyshon@astro.cf.ac.uk}}
\date{}
\end{opening}
\runningauthor{G. Leyshon}
\runningtitle{Reduction and presentation of photopolarimetry}

\newcommand{\ha}{{\frac{1}{2}}}

\newcommand{\util}[3]{{{#1}^{{#2}}_{{\sim}}{#3}}}

\newcommand{\gapeq}[2]{\util{#1\:}{>}{\:#2}}

\newcommand{\lapeq}[2]{\util{#1\:}{<}{\:#2}}

\newcommand{\bi}[1]{{\bf{#1}}}

\newcommand{\dg}{{^{\circ}}}

\newcommand{\bllt}{${\bullet}\hspace{1em}$}

\newcommand{\mL}[1]{{\mathcal{L}}(#1)}

\newcommand{\mU}[1]{{\mathcal{U}}(#1)}

\newcommand{\mA}{{\mathcal{A}}}

\newcommand{\mD}{{\mathcal{D}}}

\newcommand{\abs}[1]{{|#1|}}

\newcommand{\ah}[1]{{{\hat{a}}_{\mathsf{#1}}}}

\newcommand{\mh}[1]{{{\underline{m}}_{\mathsf{#1}}}}

\newcommand{\ml}[1]{{{m}_{\mathsf{#1}}}}

\newcommand{\eps}[1]{{{\mathcal E}_{\mathsf{#1}}}}

\newcommand{\vas}[1]{{\varepsilon_{\mathsf{#1}}}}

\newcommand{\sbb}[1]{{\hat{\sigma}_{#1}}}

\newcommand{\cls}[1]{{\bar{#1}_{C_{#1}}}}

\newcommand{\cln}[2]{{\bar{#1}_{#2\%}}}

\begin{document}
\maketitle

\begin{abstract} Many different methods exist for reducing data obtained
when an astronomical source is studied with a two-channel polarimeter,
such as a Wollaston prism system. This paper presents a rigorous method of
reducing the data from raw aperture photometry, and evaluates errors both
by a statistical treatment, and by propagating the measured sky noise from
each frame. The reduction process performs a hypothesis test for the
presence of linear polarization. The probability of there being a non-zero
polarization is obtained, and the best method of obtaining the normalized
Stokes Parameters is discussed. Point and interval estimates are obtained
for the degree of linear polarization, which is subject to positive bias; 
and the polarization axis is found. \end{abstract}

\keywords{polarization, methods: data analysis, techniques: polarimetric}

\section{Introduction}

 When performing optical polarimetry of astronomical objects, we wish to
answer three distinct, but related, physical questions.

 Firstly, is the object polarized at all? Secondly, if it is, what is the
best estimate of the polarization? And thirdly, what confidence can we
give to this measure of polarization? 

 In addition to these physical questions is a presentational one: in what 
format should the results be published, so that they will be of most 
utility to the scientific community?

 The questions of quantifying and presenting data on linear polarization 
have been discussed 
at length by Simmons \& Stewart~\shortcite{sims}, who note that the 
traditional method used by optical astronomers, that of 
Serkowski~\shortcite{serk-trad}, does not give the best estimate of the 
true polarization under most circumstances. Using their recommendations, 
I present here a recipe for reducing polarimetric data.

\section{Paradigm}

 In this paper, I will not consider the origin of the polarization of 
light. It may arise from intrinsic polarization of the source, from 
interaction with the interstellar medium, or within Earth's atmosphere. 
Each of these sources represents a genuine polarization, which must be taken 
into account in explaining the measured polarization values.

 Most modern optical polarimetry systems employ a two-channel system, 
normally a Wollaston prism. Such a prism splits the incoming light 
into two parallel beams (`channels') with orthogonal polarizations - it 
functions as a pair of co-located linear analyzers. The transmission 
axes of the analyzers can be 
changed either by placing a half-wave plate before the prism in the 
optical path, and rotating this, or by rotating the actual Wollaston 
prism. Such a system is incapable of distinguishing circularly 
polarized light from unpolarized light, and references to `unpolarized' 
light in the remainder of this paper strictly refer to light which is 
not linearly polarized; it may be totally unpolarized (i.e. randomly 
polarized), or may include a circularly polarized component.

 Where a half-wave plate is used, an anticlockwise rotation $\chi$ of the 
waveplate results in an anticlockwise rotation of $\eta = 2\chi$ of the 
transmission axes. (For the theory of Wollaston prisms and wave plates, see, 
for instance, Chapter 8 in Hecht \shortcite{Hecht}.\,)

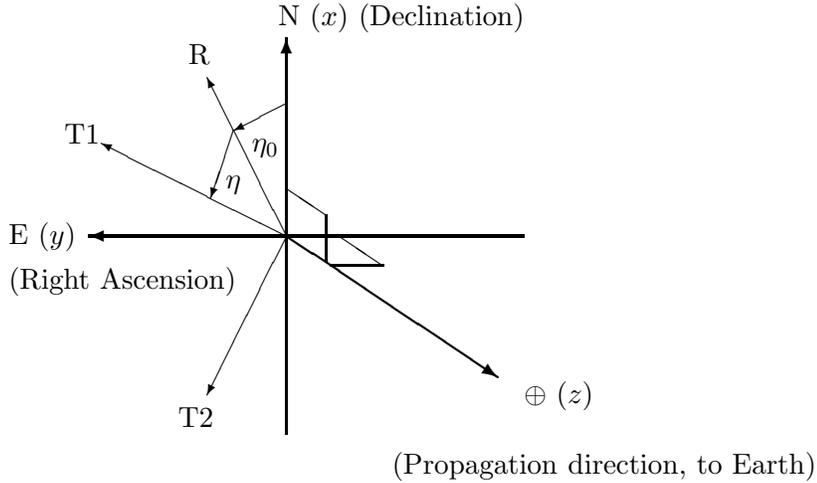
\begin{figure}
\begin{picture}(250,200)(-120,-100)
\thicklines
\put (0,-75){\vector(0,1){150}}
\put (-3,80){N $(x)$ (Declination)}
\put (90,0){\vector(-1,0){165}}
\put (-105,-3){E $(y)$}
\put (-105,-20){(Right Ascension)}
\put (0,0){\vector(3,-2){80}}
\put (90,-63){$\oplus\ (z)$}
\put (40,-90){(Propagation direction, to Earth)}
\thinlines
\put (20,0){\line(3,-2){16.64}}
\put (36.64,-11.09){\line(-1,0){20}}
\put (15,-10){\line(0,1){18.03}}
\put (15,8.03){\line(-3,2){15}}
\put (0,0){\vector(-1,2){30}}
\put (-37,65){R}
\put (-13,32){$\eta_0$}
\put (0,50){\vector(-2,-1){20}}
\put (-23,18){$\eta$}
\put (-20,40){\vector(-1,-3){8.57}}
\put (0,0){\vector(-2,1){70}}
\put (-84,35){T1}
\put (0,0){\vector(-1,-2){30}}
\put (-41,-72){T2}
\end{picture}
\caption{Reference axis, R, relative to celestial co-ordinates.}
\label{figeta}
\end{figure}

 We will suppose that Channel 1 of the 
detector has a transmission axis which can be rotated by some angle 
$\eta$ anticlockwise on the celestial sphere, relative to a reference 
position $\eta_0$ east of north. (See Figure~\ref{figeta}.)
The transmission axes T1, T2, of Channels 1 and 
2 are hence at $\eta_0 + \eta$ and $\eta_0 + 90\dg + \eta$ respectively.

 The reference angle $\eta_0$ will depend on the construction of the
polarizer, and will not, in general, be neatly due north. For mathematical
convenience in the rest of this paper, we will take $\eta_0$ to define a
reference direction, `R', in our instrumental co-ordinate system and
relate all other angles to it. Such instrumental angles can then be mapped
on to the Celestial Sphere by the addition of $\eta_0$. 

Since the light emerging in the two 
beams has traversed identical paths until reaching the 
Wollaston prism, this method of polarimetry does not suffer from
the systematic errors due to sky fluctuation which affect single-channel 
polarimetry (where a single beam polarimeter alternately samples the 
two orthogonal polarizations).

 The two channels will each feed some sort of photometric array, e.g. a {\sc 
ccd}, which will record a photon count. Since such images are usually 
built up by a process of shifting the image position on the array and 
combining the results, we will refer to a composite image taken in one 
transmission axis orientation, $\eta$, as a {\em mosaic}. We 
will denote the rate of arrival of photons recorded in Channel 1 and
Channel 2 by $n_{1}(\eta)$ and $n_{2}(\eta)$ respectively. From these
rates, we can calculate the total intensity ($I$) of the source, and
the difference ($S$) between the two channels: 

\begin{equation} 
I(\eta) = n_{1}(\eta) + n_{2}(\eta),
\label{Idef}
\end{equation}

\begin{equation} 
S(\eta) = n_{1}(\eta) - n_{2}(\eta).
\label{Sdef}
\end{equation}

We can also define a {\em normalized} difference:
\begin{equation}
\label{normdiff}
s(\eta) = \frac{S(\eta)}{I(\eta)}.
\end{equation}

The purpose of this paper is to discuss how to interpret and present such 
data.

\section{Curve Fitting for $p$}

 Suppose we have a beam of light, which has a linearly polarized component
of intensity $I_p$, whose electric vector points at an angle $\phi$
anticlockwise of R. Its unpolarized component 
is of intensity $I_u$. 
When such a beam enters our detector, we can use Malus' Law~\cite[\S
8.2.1]{Hecht} to deduce that
\[n_1(\eta) = \ha I_u + I_p.\cos^2(\phi - \eta)\]
and
\[n_2(\eta)  = \ha I_u + I_p.\sin^2(\phi - \eta),\]
from which we find
\begin{equation}
\label{Ipol}
I(\eta) = I_u + I_p,
\end{equation}
and, less trivially,
\begin{equation}
	S(\eta) = I_p.\cos[2(\phi - \eta)].
\label{Spol}
\end{equation}

The {\em degree of linear polarization}, $p$, is defined by
\begin{equation}
\label{ppol}
 p = \frac{I_p}{I_p + I_u}
\end{equation}
and so we can obtain the normalized difference by substituting Equations 
\ref{Ipol}, \ref{Spol} and \ref{ppol} into \ref{normdiff}:
 \[ s(\eta) = p.\cos[2(\phi - \eta)].\]

 Now, if observations have been made at a number of different angles,
$\eta_j$, of the transmission axis, then a series of values for $\eta_j$
and $s_j(\eta_j)$ will be known, and $p$ and $\phi$ may be determined by
fitting a sine curve to this data, weighted by errors $\sigma_{s_j}(\eta_j)$
as necessary. This method has been used, for example, by di Serego Alighieri
{\em et al.}~\shortcite[\S 2]{ali}. (Their refinement of the method 
allowed for 
the correction of the $s_j(\eta_j)$ for instrumental polarization at each 
$\eta_j$, which was necessary as they were rotating the entire camera, 
their system having no half-wave plate.)

We note that if there is any systematic bias of Channel 1 compared to 
Channel 2, this will show up as an $\eta$-independent ({\sc dc}) term 
added to the sinusoidal component when $s_j(\eta_j)$ is fitted to the data. 
Such bias could arise if an object appears close to the edge of the {\sc 
ccd} in one channel, for example.

\section{The Stokes Parameters}

 Polarized light is normally quantified using Stokes' parameterisation. 
(For basic definitions see, for example, 
Clarke, in Gehrels (ed.)~\shortcite{clarke:def}.)
Four variables are used, but one, $V$, is only applicable to circular 
polarization, which a system involving only half-wave plates and linear 
analyzers cannot measure. The total intensity, $I$, of the light is an 
absolute Stokes Parameter. The other two parameters are defined relative 
to some reference axis, which in our case will be R, the $\eta_0$ 
direction. Thus we define:
\[Q = S\,(0\dg) = - S\,(90\dg),\] 
and
\[U = S\,(45\dg) = - S\,(135\dg).\]

{\em Normalized\,} Stokes' Parameters are denoted by lower case letters 
($q$,$u$,$v$), and are found by dividing the raw parameters by $I$. We 
note that $S$ and the normalized $s$ can be thought of as a Stokes 
Parameter like $Q$ or $U$, generalised to an arbitrary angle - and 
results which can be derived for $S$ (or $s$) will apply to $Q$ and $U$ 
(or $q$ and $u$) as special cases.

 If the Stokes Parameters are known, then the degree and angle of 
polarization can be found:
\begin{equation}
\label{defp}
 p = \sqrt{q^2 + u^2};
\end{equation}
\begin{equation}
\label{phidef}
\phi = \ha.\tan^{-1}(u/q),
\end{equation}
where the signs of $q$ and $u$ must be inspected to determine the correct
quadrant for the inverse tangent. Note that $S\,(\eta)$, $Q$ and $U$ {\em
must} be defined as above to be consistent with the choice of R as
Reference. 

 We must now distinguish between the true values of the Stokes Parameters 
for a source, and the values which we measure in the presence of noise. 
We will use the subscript $0$ to denote the underlying values, and 
the subscript $i$ for individual measured values.

 In particular, consider a source which is not polarized, so $q_0=u_0=0$, 
$p_0=0$, and $\phi_0$ is undefined. Since the $q_i$ and $u_i$ include noise, 
they will not, in general, be zero, and because of the form of Equation 
\ref{defp}, $p_i$ will be a definite-positive 
quantity. In short, $p_i$ is a {\em biased} estimator for $p_0$.

 There is no known {\em unbiased} estimator for $p_0$, and Simmons \&
Stewart~\shortcite{sims} discuss at length the question of which 
estimator
should be used. They conclude that the Stokes Parameters themselves are
more useful than $p$ and $\phi$ in many applications, and it is recommended,
therefore, that all published polarimetric data should ideally give the
normalized Stokes Parameters, with or without evaluation and discussion of
$p$ and $\phi$. 

 Given this preference for the Stokes Parameters it appears that one should
eschew the curve fitting method in favour of direct evaluation of the
parameters, at least when we only have data for the usual angles $\eta_j =
0\dg, 45\dg, 90\dg, 135\dg$. In practice, observers will take several
observations of an object at each transmission angle. This raises the
question of how best to combine all the measured values $q_i, u_i$ to
yield a single pair of `best estimators' for $q_0$ and $u_0$ -- a question
which is dealt with by Clarke {\em et al.}~\shortcite{clarke}

 On the basis of this prior work and set of recommendations, it is now 
possible to present a `recipe' for reducing polarimetric data.

\section{Handling the Raw Data}

\subsection{SKY NOISE AND SHOT NOISE}

The raw numbers which our photometric system produces will be a set of
photon count rates $n_{1i}(\eta)$ and
$n_{2i}(\eta)$, together with their errors, 
$\sigma_{n_{1i}}(\eta)$
and $\sigma_{n_{2i}}(\eta)$. Errors arise from three sources: photon shot 
noise; pixel-to-pixel variations in the sky value superimposed on the target 
object; 
and imperfect estimation of the modal sky value to subtract from the 
image \cite{photbook,IRAFphot}.

 The photons arrive at the detector according to a Poisson
distribution. Let the total
integration time for a mosaic taken at a given rotation of the 
polarizer, be $\tau$. If the detector requires $e$ photons to arrive
in order for one `count' to be registered, then the total number of photons 
incident to produce the measured signal is $ne\tau$.

Under Poisson statistics, using units of `numbers of photons', the standard 
deviation of the number of photons arriving in this time-bin is the square 
root of the mean number arriving, {\em viz.}\ $\sigma_{\gamma} = 
\sqrt{ne\tau}$. In our detector-based count rate units, therefore, the error 
contributed 
is $\sigma_{\mathsf{shot}} = \sqrt{n/e\tau}$. Provided $ne\tau > 
10$,~\cite[\S 19.10]{statstext} 
then the shot noise will be normally distributed, to a good approximation.

 The modal value of a sky pixel, $n_{\mathsf{sky}}$ can be found by 
considering, say, the pixel values in an annulus of dark sky around the 
object in question, an annulus which contains $\mD$ pixels 
altogether.
 The root-mean-square deviation of these pixels' values about the mode can 
also be found, and we will label this, $\sigma_{\mathsf{sky}}$. Hence we can
estimate the error on the mode, 
$\sigma_{\mathsf{sky}}/\sqrt{\mD}$.

 If we perform aperture-limited photometry on our target, with an 
aperture of area $\mA$ in pixels, we must subtract the modal sky 
level, $\mA.n_{\mathsf{sky}}$, which will 
introduce an 
error $\sigma_{\mathsf{skysub}} = 
\mA.\sigma_{\mathsf{sky}}/\sqrt{\mD}$.

 Each individual pixel in the aperture will be subject to a random sky 
fluctuation; adding these in quadrature for each of the $\mA$ 
pixels, we obtain an error $\sigma_{\mathsf{skyfluc}} =
\sqrt{\mA}.\sigma_{\mathsf{sky}}$.

 Ultimately, the error on the measured, normalized, intensity, is the sum 
in quadrature of the three quantities, $\sigma_{\mathsf{shot}}$, 
$\sigma_{\mathsf{skysub}}$, and $\sigma_{\mathsf{skyfluc}}$. If the areas 
of the aperture and annulus are comparable, then both the second and 
third terms will be significant; in practice, for long exposure times, 
the first (shot) noise term will be much smaller and can be neglected. 
This is important as, unlike the sky noise, the shot noise depends on the 
magnitude of the target object itself. If its contribution to the error 
terms is negligible, then sky-dominated error terms can be compared 
between objects of different brightness on the same frame.

\newtheorem{step}{Data Reduction Step}

\newtheorem{chek}[step]{Data Check}

\begin{chek}
\label{smallshot}

 For each object observed in each channel of each mosaic, the photometry 
system will have produced a count rate $n_i$ 
with an error, $\sigma_{n_i}$. For each such measurement, calculate  
$\sqrt{n_i/e\tau}$ and verify that
it is much less than $\sigma_{n_i}$. Then one can be certain that the 
noise terms are dominated by sky noise rather than shot noise.

\end{chek}

\subsection{TESTING FOR dc BIAS}

In practice, for each target object, we will have taken a number of 
mosaics at each angle $\eta_j$. We can immediately use each pair of 
intensities $n_{1i}, n_{2i}$ to find $I_{i}(\eta_j)$ 
and $S_{i}(\eta_j)$ using Equations \ref{Idef} and \ref{Sdef}. 

 Since the errors 
on the two channels are independent, we can trivially find the errors on 
both $I_{i}(\eta_j)$ and $S_{i}(\eta_j)$; the errors turn out to be 
identical, and are given by:
\begin{equation}
\label{erreq}
\sigma_{I_{i}} = \sigma_{S_{i}} = \sqrt{{\sigma_{n_{1i}}}^2 + 
{\sigma_{n_{2i}}}^2 }.
\end{equation}

\vspace{1ex}

\begin{chek} 
\label{cbias}
Take the mean value of all the $S_{i}(\eta_j)$ by summing over all the 
values $S_i$ at all angles $\eta_j$; and obtain an
error on this mean by combining in quadrature the error on each $S_i$. If
the mean value of $S_i(\eta_j)$, averaged over all the angles $\eta_j$, is
significantly greater than the propagated error, then there may be some
{\sc dc} bias.
\end{chek}

 Check \ref{cbias} uses $S_{i}(\eta_j)$ as a measure of excess intensity
in Channel 1 over Channel 2, and relies on the fact that there are 
similar numbers of observations at $\eta_j=\eta$ and 
$\eta_j=\eta+90\dg$ to average away effects due to polarization. If, as 
may happen in real data gathering exercises, there are not {\em 
identical} numbers of observations at $\eta_j=\eta$ and
$\eta_j=\eta+90\dg$, this could show up as apparent `{\sc dc} bias' in a 
highly polarized object. In practice, however, we are unlikely to 
encounter this combination of events; testing for bias by the above 
method will either reveal a bias much greater than the error (where the 
cause should be obvious when the original sky images are examined); or a 
bias consistent with the random sky noise, in which case we can assume 
that there is no significant bias. 

\subsection{OBTAINING THE STOKES PARAMETERS}

 Once we are satisfied that our raw data are not biased, we can proceed.
At this stage in our data reduction, we will find it convenient to 
divide our set of $S_{i}(\eta_j)$ values, together with their associated 
$I_{i}(\eta_j)$ values, into the named Stokes Parameters,
\[Q_i = S_{i}(\eta_j=0\dg) = -S_{i}(\eta_j=90\dg)\]
and 
\[U_i = S_{i}(\eta_j=45\dg) = -S_{i}(\eta_j=135\dg).\]

 In the rest of this paper, symbols such as $S_i$ and $\sigma_{S_i}$, where 
not followed by $(\eta)$, can be read as denoting `either $Q_i$ or 
$U_i$', `either $\sigma_{Q_i}$ or $\sigma_{U_i}$', etc..

\begin{step}
\label{getthei}

For each pair of data $n_{1i}(\eta_j), n_{2i}(\eta_j)$, produce the sum,
$I_{i}$, and the difference, $Q_{i}$ or $U_{i}$ as appropriate. 
Using Equation \ref{erreq}, produce the error common to the sum and
difference, $\sigma_{Q_i}$ or $\sigma_{U_i}$. Also find the normalized 
difference, $q_i$ or $u_i$.

\end{step} 

In practice, for a given target object, we will have taken a
small number of measurements of $Q_i$ and $U_i$ -- say $\nu_Q$ and 
$\nu_U$ respectively -- with individual
errors obtained for each measurement. If the errors on the individual
values are not comparable, but vary widely, we may need to consider taking a 
weighted mean.

\begin{chek} 
\label{maxbig} 
For a set of measurements of $(S_i,\sigma_{S_i})$, take all the measured 
errors,  $\sigma_{S_i}$; and so find the mean error (call this $\eps{phot}$) 
and the maximum deviation of any individual error
from $\eps{phot}$. If the maximum deviation is large 
compared to 
the actual error, consider whether you need to weight the data. 
\end{chek}

 If the deviations are large, we can weight each data point, $S_i$, by 
${\sigma_{S_i}}^{-2}$; but we will not pursue the subject of statistical 
tests on weighted means here. In practice, one normally finds that the
noise does not vary widely between measurements.

 We have already checked (see Check \ref{smallshot}) that the shot noise is 
negligible compared with the sky noise terms. Therefore, the main source 
of variation will be the sky noise. If the maximum deviation of the 
errors from $\eps{phot}$ is small, then we can infer that the 
fluctuation in the sky pixel values is similar in all the mosaics.

\begin{step}
\label{assumenorm}
In order to carry the statistical treatment further, we must assume 
that 
the sky noise is normally distributed. This is standard astronomical 
practice.
\end{step}

\begin{step}
\label{getmean}

From the sample of Stokes Parameters $I_i$, $Q_i$ and $U_i$, obtained in 
Step \ref{getthei}, find 
the two means, $\bar{Q}$ and $\bar{U}$, with their corresponding intensities 
$\bar{I}_Q$ and $\bar{I}_U$; and find
the standard deviations of the two \bi{samples}, $\psi_Q$ and $\psi_U$.

\end{step}

\subsection{PHOTOMETRIC AND STATISTICAL ERRORS}

Since modern photometric systems can estimate the sky noise on each 
frame, we are faced throughout our data reduction sequence with a choice 
between two methods for handling errors. We can propagate the errors on 
individual measurements through our calculations; or we can use the 
standard deviation, $\psi_S$, of the set of sample values, $S_i$.

 In this paper, I use the symbol $\sigma_{S_i}$ to denote the measured 
(sky-dominated) error on $S_i$, and $\sigma_{\bar{S}}$ for the standard 
error on the estimated mean, $\bar{S}$. The standard deviation of the 
population, which is the expected error on a single measurement $S_i$, 
could be denoted $\sigma_{S}$, but above I used $\eps{phot}$ to make its 
photometric derivation obvious.

 Using statistical estimators discards the data present in the 
photometric noise figures and uses only the spread in the data points to 
estimate the errors. We would expect the statistical estimator to be of 
similar magnitude to the photometric error in each case; and a cautious 
approach will embrace the greater of the two errors as the better error 
to quote in each case.

 Because we may be dealing with a small sample (size $\nu_S$) for some 
Stokes Parameter, $S$, the standard deviation 
of the sample, $\psi_S$, will not be the best estimator of the 
population standard deviation. The best estimator is~\cite[\S 10.5, for 
example]{statstext}:
\begin{equation}
\label{stateq}
\eps{stat} = \sqrt{\frac{\nu_S}{\nu_S - 1}}.\psi_S. 
\end{equation}

 In this special case of the \bi{population} standard deviation, I have used 
the notation $\eps{stat}$ for clarity. Conventionally, $s$ is used for the 
`best estimator' standard deviation, but this symbol is already in use here 
for a general normalized Stokes Parameter, so in this paper I will use 
the variant form of sigma, $\varsigma$, for errors derived from the 
sample standard deviation, whence $\varsigma_S = \eps{stat}$, and the 
(statistical) standard error on the mean is
\[ \varsigma_{\bar{S}} = \frac{\psi_S}{\sqrt{\nu_S - 1}} = 
\frac{\eps{stat}}{\sqrt{\nu_S}}.\]

 The mean value of our Stokes Parameter, $\bar{S}$, is the best estimate 
of the true value $(S_0)$ regardless of the size of $\nu_S$. Given a 
choice of errors between $\sigma_{\bar{S}}$ and  $\varsigma_{\bar{S}}$, 
we will cautiously take the greater of the two to be the `best' error, 
which we shall denote $\sbb{\bar{S}}$.

\begin{chek}
\label{noiseOK}

We now have two ways of estimating the noise on a single 
measurement of a Stokes Parameter: 

\bllt $\eps{phot}$ is the 
mean sky noise level obtained from our photometry system: Check 
\ref{maxbig} obtains its value and verifies that the noise levels do not 
fluctuate greatly about this mean.

\bllt Statistical fluctuations in the 
actual values of the Stokes Parameter in question are quantified by 
$\eps{stat}$, obtained by applying Equation \ref{stateq} to the data 
from Step \ref{getmean}.

 We would expect the 
two noise figures to be comparable, and this can be checked in our data. 
We may also consider photometry of other objects on the same frame: 
Check \ref{smallshot} shows us that the errors are dominated by sky 
noise, and $\sigma_{\mathsf{sky}}$ should be comparable between objects, 
correcting for the different apertures used: 
\[ \sigma_{\mathsf{sky}} = \eps{X}/\sqrt{2\mA(1+\mA/\mD)}.\]

 We therefore take the best error, $\sbb{S}$, on a Stokes 
Parameter, $S$, to be the greater of $\eps{phot}$ and $\eps{stat}$.

\end{chek}

 If our data passes the above test, then we can be reasonably 
confident that the statistical tests we will outline in the next 
sections will not be invalidated by noise fluctuations.

\section{Testing for Polarization}

 The linear polarization of light can be thought of as a vector of length 
$p_0$ and
phase angle $\theta_0 = 2\phi_0$. There are two independent components to
the polarization. If either $Q_0$ or $U_0$ is non-zero, the light is said 
to be polarized. Conversely, if the light is to be described as 
unpolarized, both $Q_0$ and $U_0$ must be shown to be zero.

 The simplest way to test whether or not our target object emits polarized 
light is to test whether the measured Stokes Parameters, $\bar{Q}$ and 
$\bar{U}$, are consistent with zero. If either parameter is 
inconsistent with zero, then the source can be said to be polarized.

 To proceed, we must rely on our assumption (Step \ref{assumenorm}) that the 
sky-dominated noise
causes the raw Stokes Parameters, $Q_i, U_i,$ to be distributed normally. 
Then we can perform hypothesis testing~\cite[Chapters 12 and
16]{statstext} for the null hypotheses that $Q_0$ and $U_0$ are zero. 
Here, noting that the number of samples is typically small ($\nu_Q \simeq 
\nu_U < 30$) we face a choice: 

\vspace{1ex}

 \bllt {\sf Either:} assume that the sky fluctuations are normally 
distributed with standard deviation $\eps{phot}$, and perform 
hypothesis testing on the standard normal distribution with the statistic:
\[ z = \frac{\bar{S} - S_0}{\eps{phot}/\sqrt{\nu_S}}; \] 

 \bllt {\sf Or:} use the variation in the $S_i$ values to estimate the 
population standard deviation $\eps{stat}$, and perform 
hypothesis testing on the Student's $t$ distribution with $\nu_S - 1$ 
degrees of freedom, using the statistic:
\[ t = \frac{\bar{S} - S_0}{\eps{stat}/\sqrt{\nu_S}}.\] 

 In either case, we can perform the usual statistical test to determine
whether we can reject the null hypothesis that `$S_0 = 0$', at the
$C_S.100\%$ confidence level. The confidence intervals for retaining the
null hypothesis will be symmetrical, and will be of the forms $-z_0<z<z_0$
and $-t_0<t<t_0$.

 The values of $z_0$ and $t_0$ can be obtained from tables, and we define
$\cls{S}$ to be the greater of
$z_0.\eps{phot}/\sqrt{\nu_S}$ and
$t_0.\eps{stat}/\sqrt{\nu_S}$. Then the more conservative hypothesis test 
will reject that null hypothesis at the $C_S.100\%$ confidence level when
$\abs{\bar{S}}>\cls{S}$.

 In such a confidence test, the probability of making a `Type I Error',
i.e.\ of identifying an \bi{unpolarized} target as being polarized in {\em 
one} polarization sense, is simply $1-C_S$. The probability of correctly 
retaining the `unpolarized' hypothesis is $C_S$.

 The probability of making a `Type II Error'~\cite[\S 12.7]{statstext}
(i.e.\ not identifying a \bi{polarized} target as being polarized in one
polarization sense) is not trivial to calculate. 

Now because there are two independent senses of linear polarization, we 
must consider how to combine the results of tests on the two independent 
Stokes Parameters. Suppose we have a source which has no linear 
polarization. We test the two Stokes Parameters, $\bar{Q}$ and 
$\bar{U}$, for consistency with zero at confidence levels $C_Q$ and 
$C_U$ respectively. The combined probability of correctly retaining the 
null hypothesis for both channels is $C_Q.C_U$, and that of making the 
Type I Error of rejecting the null hypothesis in either or both channels 
is $1-C_Q.C_U$. Hence the overall confidence of the combined test is 
$C_Q.C_U.100\%$.

 Since the null hypothesis is that
$p_0=0$ and $\phi_0$ is undefined, there is no preferred direction in the 
null system, and therefore the confidence test should not prefer one 
channel over the other. Hence the test must always take place with 
$C_Q=C_U$.

 Even so, the test does not treat all angles equally; the 
probability of a Type II Error depends on the orientation of the 
polarization of the source. Clearly if its polarization is closely 
aligned with a transmission axis, there is a low chance of a polarization 
consistent with the null hypothesis being recorded on the aligned axis, 
but a much higher chance of this happening on the perpendicular axis. As 
the alignment worsens, changing $\phi_0$ while keeping $p_0$ constant, the 
probabilities for retaining the null hypothesis on the two measurement axes 
approach one another.

 Consider the case where we have taken equal numbers of measurements in
the two channels, so $\nu_Q = \nu_U = \nu$, and where the errors on the
measurements are all of order $\eps{phot}$. Hence we can calculate $z_0$
for the null hypothesis as above. Its value will be common to the $Q$ and
$U$ channels, as the noise level and the number of measurements are the
same in both channels. 

 Now suppose that the source has 
intensity $I_0$ and a true non-zero polarization $p_0$ oriented at position 
angle 
$\phi_0$. Then we can write $Q_0 = I_0 p_0 \cos(2\phi_0)$, and $U_0 = I_0 
p_0 \sin(2\phi_0)$. To generate a Type II error, a false null result 
must be recorded on both axes. The probability of a false null can be 
calculated for specified $p_0$ and $\phi_0$: defining $z_1 = 
\frac{I_0 p_0}{\eps{phot}/\sqrt{\nu}}$ then the probability of such a 
Type II error is
\begin{equation}
\label{IIprob}
P_{\rm II} = \frac{1}{2\pi}
\int_{x= z_1 \cos(2\phi_0) - z_0}^{x= z_1 \cos(2\phi_0) + z_0} 
\int_{y= z_1 \sin(2\phi_0) - z_0}^{y= z_1 \sin(2\phi_0) + z_0} 
\exp\left[ - \ha (x^2 + y^2) \right] dx\,dy.
\end{equation}
Clearly this probability is 
not independent of $\phi_0$.

\begin{step}
Find the 90\%
confidence region limits, $\cln{Q}{90}$ and $\cln{U}{90}$, and inspect whether
$\abs{\bar{Q}}<\cln{Q}{90}$ and $\abs{\bar{U}}<\cln{U}{90}$.

\bllt If both Stokes 
Parameters fall within the limits, then the target is not shown to be 
polarized at 
the 81\% confidence level. In this case we can try to find polarization 
with some lower confidence, so repeat the test for 
$C_Q=C_U=85\%$. If 
the null hypothesis can be rejected in either channel, then 
we have a detection at the 72.25\% confidence level. There
is probably little merit in plumbing lower confidences than this.

\bllt If, however, polarization is detected in one or both of the Stokes
Parameters at the starting point of 90\%, test the polarized parameters to
see if the polarization remains at higher confidences, say 95\% and
97.5\%. The highest confidence with which we can reject the null
(unpolarized) hypothesis for either Stokes Parameter should be squared to
give the confidence with which we may claim to have detected an overall
polarization. 

\end{step}

 In our hypothesis testing, we have made the {\em a priori}\ assumption 
that all targets are to be assumed unpolarized until proven otherwise. 
This is a useful question, as we must ask whether our data are worth 
processing further -- and we ask it using the raw Stokes 
Parameters, without resorting to complicated formulae. To publish useful 
results, however, we must produce the normalized Stokes Parameters, 
together with some sort of error estimate, and it is this matter which we 
will consider next.

\section{The Normalized Stokes Parameters}

 Consider a general normalized Stokes Parameter for some angle, $\eta$:
\[ s_i = \frac{S_i}{I_i} = 
\frac{n_{1i} - n_{2i}}{n_{1i} + n_{2i}}.\]
Clarke {\em et al.}~\shortcite{clarke} point out that the signal/noise ratio 
obtained by calculating 
\begin{equation}
\label{stilde}
 \tilde{s} = \frac{\bar{S}}{\bar{I}} = 
\frac{\sum_{i=1}^{\nu_S} S_i}{\sum_{i=1}^{\nu_S} I_i}
\end{equation}
is much better than that obtained by simply taking the mean,
\begin{equation}
\label{sbar}
\bar{s} = \frac{1}{\nu_s} \sum_{i=1}^{\nu_S} s_i
= \frac{1}{\nu_s} \sum_{i=1}^{\nu_S} \frac{S_i}{I_i},
\end{equation}
 since the Equation \ref{stilde} involves the taking of only one 
ratio, where the two terms $\bar{S}$ and $\bar{I}$ have better 
signal/noise ratios than the individual $S_i$ and $I_i$ which are ratioed 
in Equation \ref{sbar}.

We also note that errors on $\bar{S}$ and on $\bar{I}$ are not 
independent of one another. We can write:
\begin{equation}
\label{getstilde}
\tilde{s} =  \frac{\bar{n}_{1} - 
\bar{n}_{2}}{\bar{n}_{1} + \bar{n}_{2}}.
\end{equation}

If we propagate through the errors on the intensities, we find:
\begin{equation}
\label{getsterr}
\sigma_{\tilde{s}} =
\frac{1}{\bar{n}_{1}+\bar{n}_{2}}.\sqrt{[(1-\tilde{s})\sigma_{\bar{n}_{1}}]^2
+ [(1+\tilde{s})\sigma_{\bar{n}_{2}}]^2}.
\end{equation}

 In order to simplify the calculation, we recall that in 
Check \ref{maxbig}, we 
checked that the errors on all the $S_i$ (and hence $I_i$) were 
similar. Thus the mean error on {\em one}\ rate in {\em one}\ channel is 
$\eps{phot}/\sqrt{2}$. Since the number of measurements 
made of $S$ is $\nu_S$, then
\[ \sigma_{\bar{n}_{1i}} \simeq \sigma_{\bar{n}_{2i}} \simeq 
\eps{phot}/\sqrt{2\nu_S}\]
and the error formula approximates to:
\begin{equation}
\label{normerr}
\sigma_{\tilde{s}} = 
\tilde{s}.\eps{phot}.\sqrt{(\bar{S}^{-2}+\bar{I}^{-2})/\nu_S}. 
\end{equation} 

In practice, we will be dealing with small polarizations, so $\bar{S} \ll 
\bar{I}$, and knowing $\tilde{s}$ from Equation \ref{stilde}, then Equation 
\ref{normerr} approximates to:
\begin{equation}
\label{simerrs}
 \sigma_{\tilde{s}} \simeq 
\frac{\tilde{s}.\eps{phot}}{\bar{S}.\sqrt{\nu_S}} = 
\frac{\eps{phot}}{\bar{I}.\sqrt{\nu_S}} 
\end{equation}

 As we had before with $\eps{stat}$ and $\eps{phot}$, so now we have a choice 
of using sky photometry or the statistics to estimate errors. The above 
method gives us the photometric error on a normalized Stokes' Parameter as 
$\vas{phot} = \eps{phot}/\bar{I} = \sigma_{\tilde{s}}.\sqrt{\nu_S}$; the 
statistical method would be to 
take the root-mean-square deviation of the measured $s_i$, obtained in 
Step \ref{getthei}, about Clarke {\em et al.}'s~\shortcite{clarke} best estimator 
value, $\tilde{s}$:
\begin{equation}
\label{Nstaterr}
\vas{stat} = \varsigma_{\tilde{s}}.\sqrt{\nu_S} =
 \frac{1}{\sqrt{\nu_S-1}}.\left[{\sum_{i=1}^{\nu_S}(s_i - 
\tilde{s})^2}\right]^{\ha} \end{equation}

\begin{step}
Following the method outlined for finding $\tilde{s}$ and 
$\sigma_{\tilde{s}}$, apply Equations \ref{stilde} and \ref{simerrs} to 
the data obtained in Step \ref{getmean} to obtain $\tilde{q}$ with 
$\sigma_{\tilde{q}}$ 
and $\tilde{u}$ with $\sigma_{\tilde{u}}$. 
\end{step}

\begin{chek}
\label{stoeq}
Using $\tilde{q}$ and $\tilde{u}$, compute $\varsigma_{\tilde{q}}$ and 
$\varsigma_{\tilde{u}}$; find $\vas{stat}$ for both 
normalized Stokes Parameters, and compare it with $\vas{phot}$ in each case.
Verify also that the errors, $\vas{X}$, on the population standard 
deviations for the two Stokes Parameters are similar -- this should 
follow from the $S$-independence of Equation
\ref{simerrs} for small $\tilde{q}$ and $\tilde{u}$.
\end{chek}

So which error should one publish as the best estimate, 
$\sbb{\tilde{s}}$, on 
our final $\tilde{s}$ --- $\sigma_{\tilde{s}}$ or $\varsigma_{\tilde{s}}$\,? 
Again, a conservative 
approach would be to take the greater of the two in each case.

\begin{step}
\label{gotnorm}
Choose the more conservative error on each normalized Stokes Parameter, 
and record the results as $\tilde{q} \pm \sbb{\tilde{q}}$ and 
$\tilde{u} \pm \sbb{\tilde{u}}$. Record also the best population standard 
deviations, $\sbb{q}$ and $\sbb{u}$. 
\end{step}

\section{The Degree of Linear Polarization}

\subsection{THE DISTRIBUTION OF THE NORMALIZED STOKES PARAMETERS}

 Having obtained estimated values for $q$ and $u$, with conservative
errors, these values -- together with the reference angle $\eta_0$ -- can
and should be published as the most convenient form of data for colleagues
to work with. It is often desired, however, to express the polarization
not in terms of $q$ and $u$, but of $p$ and $\phi$. 

 Simmons \& Stewart \shortcite{sims} discuss in detail the estimation 
of the degree of linear 
polarization. Their treatment assumes that the {\em normalized}\ Stokes 
Parameters have a normal distribution, and that 
the errors on $\tilde{q}$ and $\tilde{u}$ are similar. This latter 
condition is true for small polarizations (see Check 
\ref{stoeq}), but before we can proceed, we must test whether the 
former condition is satisfied.

If one assumes (Step \ref{assumenorm}) that $n_{1}$ and $n_{2}$ are 
normally distributed, 
one can construct, following Clarke {\em et al.}~\shortcite{clarke}, a joint 
distribution for $s$ whose parameters are 
the underlying {\em population}\ means $(n_{1_0}, n_{2_0})$ and standard 
deviations $(\sigma_1, \sigma_2)$ for the photon rates $n_{1i}$ 
and $n_{2i}$. The algebra gets a little messy here, so we define three 
parameters, $\alpha, \beta, \gamma$:

\begin{equation}
\label{paralpha}
\alpha = \ha \left[ \frac{1}{{\sigma_1}^2} + 
\frac{1}{{\sigma_2}^2} \left( \frac{1-s}{1+s} \right) ^2 \right],
\end{equation}

\begin{equation}
\label{parbeta}
\beta = \ha \left[ \frac{n_{1_0}}{{\sigma_1}^2} + 
\frac{n_{2_0}}{{\sigma_2}^2} \left( \frac{1-s}{1+s} \right) \right],
\end{equation}

\begin{equation}
\label{pargamma}
\gamma = \ha \left[ \frac{{n_{1_0}}^2}{{\sigma_1}^2} + 
\frac{{n_{2_0}}^2}{{\sigma_2}^2} \right].
\end{equation}

 Using these three equations, we can write the probability distribution 
for $s$ as:
\begin{equation}
\label{sdist}
P(s) = \frac{\beta.\exp[\frac{\beta^2}{\alpha} - 
\gamma]}{\sigma_1.\sigma_2.\sqrt{\pi.\alpha^3}.(1+s)^2}.
\end{equation}

This 
can be compared to the limiting case of the normal distribution whose mean 
$\tilde{s}_0$ and standard error $\sigma_0$ are obtained by 
propagating the underlying means $(n_{1_0}, n_{2_0})$ and standard 
deviations $(\sigma_1,\sigma_2)$ through Equations 
\ref{getstilde} and \ref{getsterr}:
\begin{equation} 
\label{snorm}
P_n(s) = 
\frac{\exp[\frac{-(s - 
\tilde{s}_0)^2}{2.{\sigma_0}^2}]}{\sigma_0.\sqrt{2\pi}}; 
\end{equation}

 We can derive an expression for the ratio $R(s) = P(s)/P_n(s)$, which 
should be close to unity if the normalized Stokes Parameter, $s$, is 
approximately normally distributed.

\begin{chek}
\label{nearnormal} ~ ~

\bllt Estimate $n_{1_0}$ and $n_{2_0}$ using Equations \ref{Idef} and 
\ref{Sdef}, and the data from Step \ref{getmean}. Estimate $\sigma_1 
\simeq \sigma_2 \simeq \sbb{S}/\sqrt{2}$, where $\sbb{S}$ is obtained 
from Check \ref{noiseOK}.

\bllt Use the values of $\tilde{s}$ and $\sbb{s}$ obtained in 
Step \ref{gotnorm} as the best estimates of $\tilde{s}_0$ and 
$\sigma_{0}$.

\bllt Hence use a computer program to 
calculate and plot 
$R(s)$ in the domain $-3\sbb{s} < s < 
+3\sbb{s}$. If R(s) is close to unity throughout this domain, 
then we may treat the normalized Stokes Parameters as being normally 
distributed.
\end{chek} 

\subsection{POINT ESTIMATION OF $p$}

 If the data passes Checks \ref{stoeq} and \ref{nearnormal}, then we can 
follow 
the method of Simmons \& Stewart~\shortcite{sims}. They `normalize' the
intensity-normalized Stokes Parameters, $q$ and $u$, by dividing them by 
their common population standard
deviation, $\sigma$. For clarity of notation, in a field where one 
can be discussing both probability and polarization, I will recast their
formulae, such that the {\em measured}\ degree of polarization, 
normalized as required, is here given in
the form $m = \tilde{p}/\sigma$; and the {\em actual}\ (underlying) degree of
polarization, also normalized, is
$a = p_0/\sigma$. It follows from the definition of $p$ (Equation 
\ref{defp}) that
\begin{equation}
\label{errp}
\sigma_p = \sqrt{\frac{q^2.{\sigma_q}^2 + u^2.{\sigma_u}^2}{q^2+u^2}}.
\end{equation}
If ${\sigma_q} = {\sigma_u} = \sigma$, then $\sigma_p = \sigma.\sqrt{2}$.

 Now, Simmons \& Stewart~\shortcite{sims} consider the case of a `single 
measurement' of each of $q$ and $u$, whereas we have found our best 
estimate of these parameters following the method of Clarke {\em et 
al.}~\shortcite{clarke} However, we can consider the whole process described 
by Clarke {\em et al.}~\shortcite{clarke} as `a measurement', and so the 
treatment holds when applied to our best estimate of the normalized Stokes 
Parameters, together with the error on that estimate.

\begin{step}
\label{findperr}
Find $\sbb{p}$, and hence $\sigma=\sbb{p}/\sqrt{2}$, by substituting our best 
estimates of $q$ and $u$ and their 
errors (Step \ref{gotnorm}) ino Equation \ref{errp}. Hence calculate $m$:
\[ m = \sqrt{\tilde{q}^2+\tilde{u}^2}/\sigma.\]
\end{step}

 The probability distribution $F(m,a)$ of obtaining a measured value, 
$m$, for some underlying value, $a$, is given by
the Rice distribution~\cite{sims,wardle}, which is cast in the current 
notation using the modified Bessel function, $I_0$~\cite[as defined in 
Ch.12, \S17]{Boas}:
 \begin{equation} 
\label{rice}
 F(m,a) = m.\exp \left[ \frac{-(a^2+m^2)}{2} \right] .I_0(ma) \ldots 
(m\geq 0)
\end{equation}
\[F(m,a)=0 {\mathit{~otherwise}}.\]

 Simmons \& Stewart \shortcite{sims} have tested various estimators 
$\ah{X}$ for bias. They find that when $\lapeq{a}{0.7}$, the best 
estimator is the `Maximum Likelihood Estimator', $\ah{ML}$, which 
maximises $F(m,a)$ with respect to $a$. So $\ah{ML}$ is the 
solution for $a$ of:
 \begin{equation}
\label{MLest}
a.I_0(ma) - m.I_1(ma) = 0.
\end{equation}
 
If $m<1.41$ then the solution of this equation is $\ah{ML} = 0$.
 
 When $\gapeq{a}{0.7}$, the best estimator is that traditionally used by
radio
as\-tron\-o\-mers, e.g.\ Wardle \& Kronberg~\shortcite{wardle}. In this case,
the best
estimator, $\ah{WK}$, is that which maximises $F(m,a)$ with respect to m, 
being the solution for $a$ of:
\begin{equation}
\label{WKest}
(1-m^2).I_0(ma) + ma.I_1(ma) = 0.
\end{equation}
If $m<1.00$ then the solution of this equation is $\ah{WK} = 0$.

 Simmons \& Stewart~\shortcite{sims} graph $m(a)$ for both cases, and so 
show 
that $m$ is a monotonically increasing function of $a$, and that $\ah{ML} < 
\ah{WK} < m$\ $\forall m$. But which estimator should one use? Under their 
treatment, the selection of one of these estimators over the other depends 
on the underlying value of $a$; they point out 
that there may be good {\em a priori}\ reasons to assume greater or 
lesser polarizations depending upon the nature of the source.

 If we do not make any such assumptions, we can use monotonicity of $m$ and 
the inequality $\ah{ML} < \ah{WK}$\ $\forall m$, to find two limiting cases:

\bllt Let $\ml{WKmin}$ be the solution of the Wardle \& Kronberg Equation 
(\ref{WKest}) 
for $m$ with $a=0.6$. Hence if $m<\ml{WKmin}$, then $\ah{ML} < \ah{WK} 
< 0.7$ and the Maximum Likelihood estimator is certainly the most 
appropriate. 
Calculating, we find $\ml{WKmin} = 1.0982 \ll 1.41$ and so the Maximum 
Likelihood estimator will in fact be zero.

\bllt Let $\ml{MLmax}$ be the solution of Maximum Likelihood Equation 
(\ref{MLest}) for $m$ with $a=0.8$. We find $\ml{MLmax} = 1.5347$.
 Hence if $m>\ml{MLmax}$, then $0.7 < \ah{ML} < 
\ah{WK}$, and Wardle \& Kronberg's estimator will clearly be the 
most appropriate.

 Between these two extremes,
we have $\lapeq{\ah{ML}}{0}.\lapeq{7}{\ah{WK}}$.
This presents a problem, in that each estimator suggests 
that its estimate is more appropriate than that of the other estimator. 
If our measured value is $\ml{WKmin} < m < \ml{MLmax}$, what should we 
take as our best estimate? We could take the mean of the two estimators, 
but this would divide the codomain of $\hat{a}(m)$ into three discontinuous 
regions; there might be some possible polarization which this method 
could never predict! It would be better, then, to interpolate between the 
two extremes, such that in the range $\ml{WKmin} < m < \ml{MLmax}$,
\begin{equation}
\label{interpa}
\hat{a} = \frac{m-\ml{WKmin}}{\ml{MLmax}-\ml{WKmin}}.\ah{ML} +  
\frac{\ml{MLmax}-m}{\ml{MLmax}-\ml{WKmin}}.\ah{WK}.
\end{equation}

If we do not know, {\em a priori}, whether a source is likely to be 
unpolarized, polarized to less than 1\%, or with a greater polarization, 
then $\hat{a}$ would seem to be a reasonable estimator of the true 
noise-normalized polarization, and certainly better than the biased $m$.

\begin{step}
\label{esta}
Use the above criteria to find $\hat{a}$, and hence obtain the best 
estimate, $\hat{p} = \hat{a}.\sigma$, of 
the true polarization of the target.
\end{step}

\subsection{A CONFIDENCE INTERVAL FOR $p$}

 As well as a point estimate for $p$, we would like error bars. The 
Rice distribution, Equation \ref{rice}, gives the probability of 
obtaining some $m$ given $a$, and can, therefore, be used to find a 
confidence 
interval for the likely values of $m$ given $a$. We can define two 
functions, $\mL{a}$ and $\mU{a}$, which give the 
lower and upper confidence limits for $m$, with some confidence $C_p$; 
integrating the Rice distribution, these will satisfy:
\begin{equation}
\label{Ldef}
\int_{m=-\infty}^{m=\mL{a}} F(m,a).dm = p_1
\end{equation}
and
\begin{equation}
\label{Udef}
\int_{m=\mU{a}}^{m=+\infty} F(m,a).dm = p_2
\end{equation}
such that 
\begin{equation}
\label{addprobs}
1 - C_p = p_1 + p_2.
\end{equation}

 Such confidence intervals are non-unique, and we need to impose an 
additional constraint. We could require that the tails outside the 
confidence region be equal, $p_1 = p_2$, but following Simmons \& 
Stewart~\shortcite{sims}, we shall require that the confidence interval have 
the smallest possible width, in which case our additional constraint is:
\begin{equation}
\label{aconst}
F[\mU{a},a] = F[\mL{a},a].
\end{equation}

\begin{figure}
\psfig{file=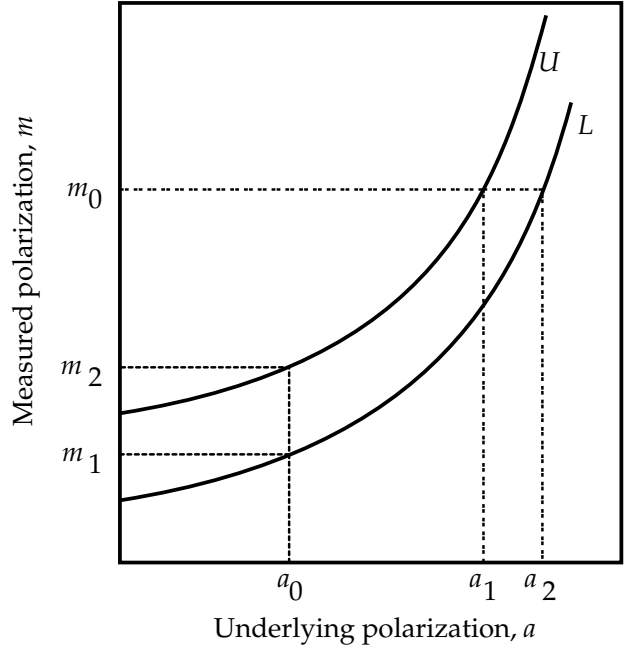,width=85mm} 
\caption{Confidence Intervals based on the Rice Distribution. Figure 
adapted from Leyshon \& Eales (1997).}
\label{ricefig} 
\end{figure}

 From the form of the Rice distribution, $\mL{a}$ 
and $\mU{a}$ will be monotonically increasing functions of $a$, as shown 
in Figure \ref{ricefig}. Given a particular underlying polarization 
$a_0$, the $C_p$ confidence interval $(m_1,m_2)$ can be 
obtained by numerically solving Equations \ref{Ldef} thru \ref{aconst} to 
yield $m_1 = \mL{a_0}$ and $m_2 = \mU{a_0}$.

 Now, it can be shown~\cite[Ch.\
VIII, \S 4.2]{moodstat} that the process can also be inverted, i.e. if we 
have obtained some measured 
value $m_0$, then solving for $m_0 = \mU{a_1} = \mL{a_2}$ will yield a
confidence interval $(a_1,a_2)$, such that the confidence of $a$ lying 
within this interval is $C_p$.

 Since the contours for $\mU{a}$ and $\mL{a}$ cut the $m$-axis at 
non-zero values of $m$, we must
distinguish three cases, depending on whether or not $m_0$ lies above one or 
both of the intercepts. The values of $\mL{0}$ and $\mU{0}$ depend only 
on the confidence interval chosen; substituting $a = 0$ into Equations 
\ref{Ldef} thru \ref{aconst} results in the pair of equations
\begin{equation}
\label{cforz}
C_m = \exp \left[ -\frac{\mL{0}^2}{2} \right] - 
      \exp \left[ -\frac{\mU{0}^2}{2} \right]
\end{equation}
and
\begin{equation}
\label{transce}
 \mL{0}.\exp \left[ -\frac{\mL{0}^2}{2} \right] =
 \mU{0}.\exp \left[ -\frac{\mU{0}^2}{2} \right].
\end{equation}

A numerical solution of this pair of equations can be found for any given 
confidence interval, $C_m$; we find that, in 67\% 
$(1\sigma)$ interval, $\mL{0} = 0.4438,\: \mU{0} = 1.6968$, while in a 
95\% $(2\sigma)$ interval, $\mL{0} = 0.1094,\: \mU{0} = 2.5048$.
Hence, knowing $m_0$, and having chosen our desired confidence level, we 
can determine the interval $(a_1,a_2)$ by the following criteria:

\bllt {\boldmath $m_0 \geq \mU{0}$} \newline
\hspace*{1.2cm}\dotfill\ There are non-zero solutions for both 
$\mU{a_1}$ and $\mL{a_2}$.

\bllt {\boldmath $\mL{0} < m_0 < \mU{0}$} \newline
\hspace*{1.2cm}\dotfill\ In this case, $a_1=0$, and we must solve $m_0 = 
\mL{a_2}$.

\bllt {\boldmath $m_0 \leq \mL{0}$} \newline
\hspace*{1.2cm}\dotfill\ Here, $a_1=a_2=0$.

 Simmons \& Stewart~\shortcite{sims} note that the third case is formally a
confidence interval of zero width, and suggest that this is
counter-intuitive; and they go on to suggest an {\em ad hoc}\ method of
obtaining a non-zero interval. However, it is
perfectly reasonable to find a finite probability that the degree of
polarization is identically zero: the source may, after all, be 
unpolarized. This can be used as the basis of estimating the probability 
that there is a non-zero underlying polarization, as will be shown in the 
next section.

\begin{step}
\label{getint}
Knowing $m$ from Step \ref{findperr},
 find the limits $(a_1,a_2)$ appropriate to confidence intervals of 67\% and
95\%. Hence, multiplying by $\sigma$, find the confidence 
intervals on the estimated degree of polarization. The 67\% limits may be 
quoted as the `error' on the best estimate.
\end{step}

\subsection{THE PROBABILITY OF THERE BEING POLARIZATION}

Consider the contour $m=\mU{a}$ on Figure \ref{ricefig}. As defined by 
Equation \ref{Udef} and the inversion of Mood {\it et 
al.}~\shortcite{moodstat}, it divides the domain into two regions, such 
that there is a probability $p_2$ of the underlying polarization being 
greater than $a={\mathcal U}^{-1}(m_0)$. There is clearly a limiting case 
where the contour cuts the $m$-axis at $m_0$, hence dividing the domain 
into the polarized region $a>0$ with probability $p_P$, and the 
unpolarized region with probability $1-p_P$.

 Now we may substitute the Rice Distribution, Equation \ref{rice}, into 
Equation \ref{Udef} and evaluate it analytically for the limiting case, 
$a=0$:
\begin{equation}
\label{propol}
p_P = 1 - \exp(-{m_0}^2/2).
\end{equation}
  Equation \ref{propol} hence yields the probability that a measured
source actually has an underlying polarization.

\begin{step}
\label{estpolun}
Substitute $m$ from Step \ref{findperr} into Equation \ref{propol}. Hence 
quote the probability that the observed source is truly polarized.
\end{step}
 
\section{The Polarization Axis}

 It remains to determine the axis of polarization, for which an unbiased
estimate is given by Equation \ref{phidef}. Once again, we
have a choice of using the statistical or photometric errors --- and,
indeed, a choice of raw or normalized Stokes Parameters. Since 
\begin{equation}
\label{redf}
2\phi = \theta = \tan^{-1}(u/q) = \tan^{-1}(r),
\end{equation}
 our first problem is to obtain the 
best figure for $r = u/q$.

 Now, as we saw in our discussion of the best normalized Stokes 
Parameter, it is better to ratio a pair of means than to take the mean 
of a set of ratios. We could take
$r=\bar{U}/\bar{Q}$, but for a very small sample, there is the danger that 
the mean intensity of the $Q$ observations will differ from that of the 
$U$ values. Therefore, we should use the normalized Stokes Parameters, 
and the least error prone estimate of the required ratio will be 
$\tilde{r}=\tilde{u}/\tilde{q}$, yielding $\tilde{\phi}$.

 Knowing the errors on $\tilde{q}$ and $\tilde{u}$, we can find the 
propagated error in $\tilde{r}$:
\begin{equation}
\label{getrerr}
\sigma_{\tilde{r}} = 
\tilde{r}.\sqrt{\left(\frac{\tilde{q}}{\sigma_{\tilde{q}}}\right)^2 + 
\left(\frac{\tilde{u}}{\sigma_{\tilde{u}}}\right)^2};
\end{equation}
given the non-linear nature of the tan function, the error on 
$\tilde{\phi}$ should be found by separately calculating $\sigma_+ = \ha 
\tan^{-1}(\tilde{r}+\sigma_{\tilde{r}}) - \tilde{\phi}$ and $\sigma_- = \ha
\tan^{-1}(\tilde{r}-\sigma_{\tilde{r}}) - \tilde{\phi}$. Careful 
attention must be paid in the case where the error takes the phase 
angle across the boundary between the first and fourth quadrants, as the 
addition of $\pm \pi$ to the inverse tangent may be neccessary to yield a 
sensible error in the phase angle.

\begin{step}
\label{propphi}
Obtain $\tilde{\phi}$, the best estimate of $\phi$, and the propagated 
error on it, $\sigma_{\tilde{\phi}} = \ha (|\sigma_+| + 
|\sigma_-|)$, using Equations \ref{redf} 
and \ref{getrerr}. Add $\eta_0$ to $\tilde{\phi}$ and hence quote the best 
estimate of the polarization orientation in true celestial co-ordinates.
\end{step}

 For the statistical error, we note that the probability distribution 
of observed {\em phase}\ angles, $\theta=2\phi$, calculated by 
Vinokur~\shortcite{vin}, and quoted in Wardle \& 
Kronberg~\shortcite{wardle}, is:
\[P(\theta) = \exp \left[ -\frac{a^2 \sin^2(\theta-\theta_0)}{2} \right] 
.\]
 \begin{equation}
\label{thetadis}
\left\{ \frac{1}{2\pi} \exp \left[ - \frac{a^2 \cos^2(\theta-\theta_0)}{2} 
\right] + \frac{a \cos(\theta-\theta_0)}{\sqrt{2\pi}}.\left\{ \ha + f[a 
\cos(\theta-\theta_0)] \right\} \right\}
\end{equation}
where
\begin{equation}
\label{thetasup}
f(x) = \frac{{\mathrm sign}(x)}{\sqrt{2\pi}} \int_0^x 
\exp \left(- \frac{z^2}{2} \right) \,dz 
={\mathrm sign}(x).{\mathrm erf}(x)/\sqrt{8}, \end{equation}
and ${\mathrm erf}(x)$ is the error function as defined in 
Boas~\shortcite{Boas}, Ch.11, \S9. We do not 
know $a=p_0/\sigma$, and will have to use our best estimate, $\hat{a}$, 
as obtained from Step \ref{esta}.
The $C_\phi.100\%$ confidence interval on the measured angle, 
$(\theta_1,\theta_2)$, is given by numerically solving
\begin{equation}
\label{thetaerr}
\int_{\theta_1}^{\theta_2}  P(\theta).d\theta = C_\phi;
\end{equation}
in this case we choose the symmetric interval, $\theta_2-\tilde{\theta} = 
\tilde{\theta}-\theta_1$.

\begin{step}
\label{findangle}
Obtain the limiting values of $\phi=\theta/2$ for confidence intervals of 
67\% $(1\sigma)$ and 95\% $(2\sigma)$. Quote the 67\% limits as 
$\varsigma_{\tilde{\phi}} = (\phi_2-\phi_1)/2$. Choose the more 
conservative error from $\varsigma_{\tilde{\phi}}$ and 
$\sigma_{\tilde{\phi}}$ as the best error, $\sbb{\tilde{\phi}}$.
\end{step}

\section{Comparison with Other Common Techniques}

 It may be instructive to note how the process of reducing polarimetric
data outlined in this paper compares with the methods commonly used in the
existing literature. The paper by Simmons \& Stewart~\shortcite{sims}
gives a thorough review of five possible point estimators for the degree
of polarisation. One of these methods is the trivial $m$ as an estimator 
of $a$. The other four methods all involve the calculation of thresholds 
$\mh{X}$: if $m < \mh{X}$ then $\ah{X}=0$. These four methods are 
the following:
\begin{enumerate}

\item Maximum Likelihood: as defined above, $\ah{ML}$ is the value of $a$
which maximises $F(m,a)$ with respect to $a$. Hence $\ah{ML}$ is the
solution for $a$ of Equation \ref{MLest}. The limit $\mh{ML}=1.41$ is 
found by a numerical method.

\item Median: $\ah{med}$ fixes the distribution of possible measured 
values such that the actual measured value is the {\em median}, hence 
$\int_{m'=0}^{m'=m} F(m',\ah{med}).dm' = 0.5$. The threshold is $\mh{med} 
= 1.18$, being the solution of $\int_{m'=0}^{m'=\mh{med}} F(m',0).dm'=0.5$.

\item Serkowski's estimator: $\ah{Serk}$ fixes the distribution of 
possible measured values such that the actual measured value is the {\em 
mean}, hence $\int_{m'=0}^{m'=\infty} m'.F(m',\ah{Serk}).dm' = m$. The 
threshold is $\mh{Serk} = 1.25 = \int_{m'=0}^{m'=\infty} m'.F(m',0).dm'$.

 \item Wardle \& Kronberg's method: as defined above, the estimator, 
$\ah{WK}$, is that which maximises $F(m,a)$ with respect to $m$ (see 
Equation \ref{WKest}), and $\mh{WK}=1.00$.

\end{enumerate}
Simmons \& Stewart~\shortcite{sims} note that although widely used in the
optical astronomy literature, Serkowski's estinator is not the best for
either high or low polarizations; they find that the Wardle \& Kronberg
method commonly used by radio astronomers is best when $\gapeq{a}{0.7}$,
i.e. when the underlying polarization is high and/or the measurement noise
is very low. The Maximum Likelihood method, superior when $\lapeq{a}{0.7}$
(i.e. in `difficult' conditions of low polarization and/or high noise),
appears to be unknown in the earlier literature.

 In this paper, I have merely provided an interpolation scheme between the
point estimators which they have shown to be appropriate to the `easy' and
`difficult' measurement regimes. The construction of a confidence interval
to estimate the error is actually independent of
the choice of point estimator, although (as mentioned above) I believe 
that Simmons \& Stewart's \shortcite[\S 3]{sims} unwillingness to `accept 
sets of zero interval as confidence intervals' is unfounded, since 
physical intuition allows for the possibility of truly unpolarised 
sources (i.e. with identically zero polarizations), and their arbitrary 
method of avoiding zero-width intervals can be dispensed with.

\section{Conclusion}

 The reduction of polarimetric data can seem a daunting task to the
neophyte in the field. In this paper, I have attempted to bring together
in one place the many recommendations made for the reduction and 
presentation of polarimetry, especially those of Simmons \&
Stewart~\shortcite{sims}, and of Clarke {\em et al.}~\shortcite{clarke}. 
In addition, I have suggested that it is possible to develop the
statistical technique used by Simmons \& Stewart~\shortcite{sims} to
obtain a simple probability that a measured object has non-zero underlying
polarization.
 I have also suggested that there is a form of estimator for 
the overall degree of linear
polarization which is more generally applicable than either the Maximum
Likelihood or the Wardle \& Kronberg~\shortcite{wardle} estimators
traditionally used, and which is especially relevant in cases where the 
measured
data include degrees of polarization of order 0.7 times the estimated
error.

 Modern computer systems can estimate the noise on each individual
mosaic of a sequence of images; this is useful information, and is not to
be discarded in favour of a crude statistical analysis. A recurring theme
in this paper has been the comparison of the errors estimated from
propagating the known sky noise, and from applying sampling theory to the
measured intensities. Bearing this in mind,
I have presented here a process for data reduction in
the form of \ref{findangle} rigorous steps and checks. The recipe might be 
used
as the basis of an automated data reduction process, and I hope that it
will be of particular use to the researcher -- automated or 
otherwise -- who is attempting polarimetry for the first time. 

\acknowledgements{I would like to thank Steve Eales for his guidance with 
this project; Bob Thomson for his advice on 
statistics; and Jim Hough, Chris Packham, Mike Disney and Mike 
Edmunds for useful references.
This work was funded by a
{\sc pparc} postgraduate research student award.}

\begin{theendnotes}
\end{theendnotes}

\begin{thebibliography}{}

\bibitem[\protect\citeauthoryear{Boas}{1983}]{Boas}
Boas, Mary L.
\newblock {\em Mathematical Methods in the Physical Sciences}.
\newblock John Wiley \& Sons, New York, second edition, 1983.

\bibitem[\protect\citeauthoryear{Clarke \& Cooke}{1983}]{statstext}
Clarke, G.~M. and Cooke, D.
\newblock {\em A Basic Course in Statistics}.
\newblock Edward Arnold, second edition, 1983.

\bibitem[\protect\citeauthoryear{Clarke et al.}{1983}]{clarke}
Clarke, D., Stewart, B.~G., Schwarz, H.~E., and Brooks, A.
\newblock The statistical behaviour of normalized stokes parameters.
\endnote{In an attempt to use more consistent notation, my paper uses
  $\bar{s}$ for the arithmetic mean of a set of parameters, $\tilde{s}$ for a
  ratio of means, and $\hat{s}$ for the best (conservative) errors on certain
  quantities. Clarke {\em et al.}, however, use $\bar{s}$ for the ratio of
  non-normalized mean Stokes Parameters, and $\tilde{s}(1)$ for the arithmetic
  mean.}
\newblock {\em Astronomy and Astrophysics}, 126:260--264, 1983.

\bibitem[\protect\citeauthoryear{di Serego Alighieri et al.}{1993}]{ali}
di~Serego~Alighieri, S., Cimatti, A., and Fosbury, R.~A.~E. 
\newblock Imaging polarimetry of high-redshift radio galaxies.
\endnote{My paper uses $\eta$ for the instrumental angle which di
  Serego Alighieri {\em et al.}\ call $\phi$.}
\newblock {\em Astrophysical Journal}, 404:584--592, 1993.

\bibitem[\protect\citeauthoryear{Hecht}{1987}]{Hecht}
Hecht, Eugene.
\newblock {\em Optics}.
\newblock Addison-Wesley, second edition, 1987.

\bibitem[\protect\citeauthoryear{Clarke, in Gehrels (ed.)}{1974}]{clarke:def}
Gehrels, Tom, editor.
\newblock {\em Planets, Stars and Nebulae studied with Photopolarimetry},
  chapter on Polarimetric Definitions (by D. Clarke), pages 45--53.
\newblock The University of Arizona Press, 1974.

\bibitem[\protect\citeauthoryear{Leyshon \& Eales}{1997}]{leyeales}
Leyshon, Gareth, and Eales, Stephen A.
\newblock K-band polarimetry of seven high-redshift radio galaxies.
\newblock {\em Monthly Notices of the Royal Astronomical 
Society}, in press, 1997. (E-print {\tt astro-ph/9708085}.)

\bibitem[\protect\citeauthoryear{Mood {\em et al.}}{1974}]{moodstat}
Mood, A.~M., Graybill, F.~A., and Boes, D.~C.
\newblock {\em Introduction to the Theory of Statistics}.
\newblock McGraw-Hill, third edition, 1974.

\bibitem[\protect\citeauthoryear{NOAO}{IRAF}]{IRAFphot}
NOAO.
\newblock {\em On-Line Documentation for IRAF: APPHOT.PHOT}.
\newblock On-line manual page for the PHOT command in the DIGIPHOT.APPHOT
  package of the IRAF data reduction system.

\bibitem[\protect\citeauthoryear{Serkowski}{1958}]{serk-trad}
Serkowski, K.
\newblock {\em Acta Astronomica}, 8:135, 1958.

\bibitem[\protect\citeauthoryear{Simmons \& Stewart}{1985}]{sims}
Simmons, J.~F.~L., and Stewart, B.~G.
\newblock Point and interval estimation of the true unbiased degree of linear
  polarization in the presence of low signal-to-noise ratios.
\newblock {\em Astronomy and Astrophysics}, 142:100--106, 1985.

\bibitem[\protect\citeauthoryear{Sterken \& Manfroid}{1992}]{photbook}
Sterken, Chr., and Manfroid, J.
\newblock {\em Astronomical Photometry: A Guide}.
\newblock Kluwer Academic, 1992.

\bibitem[\protect\citeauthoryear{Vinokur}{1965}]{vin}
Vinokur, Marc.
\newblock Optimisation dans la recherche d'une sinuso\"{\i}de de p\'{e}riode 
connue en presence de bruit. Application a la radioastronomie.
\newblock {\em Annales d'Astrophysique}, 28:412, 1965.

\bibitem[\protect\citeauthoryear{Wardle \& Kronberg}{1974}]{wardle}
Wardle, J.~F.~C., and Kronberg, P.~P..
\newblock The linear polarization of quasi-stellar radio sources at 3.71 and
  11.1 centimetres.\endnote{The Wardle \& Kronberg paper reproduces 
Vinokur's equation 
(my Equation \ref{thetadis}) but omits the factor `${\mathrm sign}(x)$' 
from Equation \ref{thetasup} on the grounds (Wardle, private 
communication) that the probability of $x$ falling in the domain $x<0$ is 
negligibly small.} \newblock {\em Astrophysical Journal}, 194:249--255, 1974.
\end{thebibliography}
\end{document}